\def\papertitle{Simulation-Based Plate-Reverb Parameter Estimation from a Single Impulse Response}
\def\paperauthorA{Minhui Lu}
\def\paperauthorB{Joshua D. Reiss}

\documentclass[twoside,a4paper]{article}
\usepackage{etoolbox}
\usepackage[taskA]{dafx26challenge}

\usepackage{amsmath,amssymb,amsfonts}
\usepackage{booktabs}
\usepackage{siunitx}
\usepackage[T1]{fontenc}
\usepackage[utf8]{inputenc}
\usepackage[english]{babel}
\usepackage{caption}
\AtBeginDocument{\urlstyle{same}}
\newcommand{\taskADemoLink}{\url{https://minhuilu.github.io/dafx26-taska-demo/}}

\input glyphtounicode
\ninept
\newcounter{numauth}
\newcounter{listcnt}
\newcommand\authcnt[1]{\ifdefined#1 \stepcounter{numauth} \fi}
\newcommand\addauth[1]{\ifdefined#1 \stepcounter{listcnt}\ifnum \value{listcnt}<\value{numauth}\appto\authorslist{, #1}\else\appto\authorslist{~and~#1}\fi\fi}
\authcnt{\paperauthorB}
\authcnt{\paperauthorC}
\authcnt{\paperauthorD}
\authcnt{\paperauthorE}
\authcnt{\paperauthorF}
\authcnt{\paperauthorG}
\authcnt{\paperauthorH}
\authcnt{\paperauthorI}
\authcnt{\paperauthorJ}
\def\authorslist{\paperauthorA}
\addauth{\paperauthorB}
\addauth{\paperauthorC}
\addauth{\paperauthorD}
\addauth{\paperauthorE}
\addauth{\paperauthorF}
\addauth{\paperauthorG}
\addauth{\paperauthorH}
\addauth{\paperauthorI}
\addauth{\paperauthorJ}

\usepackage{times}
\newif\ifpdf
\ifx\pdfoutput\relax
\else
   \ifcase\pdfoutput
      \pdffalse
   \else
      \pdftrue
   \fi
\fi

\ifpdf
  \usepackage[pdftex,
    pdftitle={\papertitle},
    pdfauthor={\authorslist},
    pdfsubject={Proceedings of the 29th International Conference on Digital Audio Effects (DAFx26)},
    colorlinks=false,
    bookmarksnumbered,
    pdfstartview=XYZ
  ]{hyperref}
  \usepackage[pdftex]{graphicx}
\else
  \usepackage[dvips]{epsfig,graphicx}
  \usepackage[dvips,
    pdftitle={\papertitle},
    pdfauthor={\authorslist},
    pdfsubject={Proceedings of the 29th International Conference on Digital Audio Effects (DAFx26)},
    colorlinks=false,
    bookmarksnumbered,
    pdfstartview=XYZ
  ]{hyperref}
\fi
\usepackage{tikz}
\usetikzlibrary{arrows.meta,positioning}

\title{\papertitle}
\affiliation
{\paperauthorA~and~\paperauthorB}
{Centre for Digital Music \\ Queen Mary University of London \\ London, United Kingdom\\
{\tt \href{mailto:minhui.lu@qmul.ac.uk}{minhui.lu@qmul.ac.uk},
\href{mailto:joshua.reiss@qmul.ac.uk}{joshua.reiss@qmul.ac.uk}}}

\begin{document}
\ifpdf
  \DeclareGraphicsExtensions{.png,.jpg,.pdf}
\else
  \DeclareGraphicsExtensions{.eps}
\fi
\maketitle

\begin{abstract}
We present a simulation-trained, non-iterative estimator for Task A of the 1st
DAFx Parameter Estimation Challenge. Each unnormalized plate-reverb impulse
response is summarized by amplitude, spectral, and decay descriptors, and an
ensemble of tree regressors estimates the six target parameters in one pass.
Across two independent synthetic validation sets, the normalized models
outperform the training-set mean and an earlier raw-regression baseline. On a
shared set, the final ensemble also outperforms a single run of the official
default PSO at substantially lower inference cost. Since the official labels
are hidden, parameter accuracy is measured on simulator-matched data, and the
released responses support only audio-side consistency checks. The estimator
returns point estimates without uncertainty.
\end{abstract}

\section{Introduction}

The 1st DAFx Parameter Estimation Challenge asks how accurately physical
audio-model parameters can be estimated from impulse responses.\footnote{
\url{https://github.com/LOGUNIVPM/1st-DAFx-Challenge}.}
The two challenge tasks operate at different levels of the same generative
process: physical and observation parameters determine a modal response, whose
damped components form the impulse response. Task A estimates the compact
physical and observation description, whereas Task B estimates the generally
much larger set of modal frequencies, decay rates, and amplitudes. This paper
addresses Task A, which requires estimating the six-dimensional parameter vector
\begin{equation}
    \theta = [\mu, D/\mu, T_0/\mu, L_y, x_o, y_o],
\end{equation}
where $\mu$ is the surface density, $D$ is the plate flexural rigidity,
$D/\mu$ and $T_0/\mu$ are the normalized stiffness and tension parameters,
$L_y$ is the plate dimension in the $y$ direction, and $(x_o,y_o)$ denotes the
output location at which the plate displacement is observed.

At inference time, each unknown parameter vector must be estimated from a
single impulse response. The parameters affect different aspects of the
response: the stiffness and tension ratios and the plate dimension shape the
modal-frequency pattern, the surface density controls the absolute response
scale, and the output location determines the relative visibility of the
modes in the measured response \cite{ren2013example,stephan2012sensor}.
These cues are distributed across a dense resonant response in which many
modes overlap in frequency and time \cite{ege2009high}. Estimation therefore
requires the overall frequency, amplitude, and temporal structure rather than a
few isolated spectral peaks.

One strategy first estimates a modal representation and then maps it to physical
parameters. Classical modal analysis represents an impulse response as damped
sinusoids and estimates their frequencies, decay rates, and amplitudes through
Hankel-matrix, ESPRIT-type, or matrix-pencil formulations
\cite{roy1989esprit,hua1990matrix}. Dense, overlapping resonances make this first
stage difficult, and the resulting modal parameters still need to be mapped to
the six Task A targets.

An alternative is analysis-by-synthesis, in which candidate physical
parameters are passed through a simulator and iteratively updated to reduce
the discrepancy between the simulated and target responses. Differentiable
signal-processing and modal-simulation frameworks make this procedure
amenable to gradient-based optimization
\cite{engel2020ddsp,jin2024diffsound,diaz2025fast}. These approaches retain
an explicit physical model, but per-instance inference requires repeated
simulation and iterative optimization for every target response. The challenge
reference baseline follows this route using particle swarm optimization
\cite{kennedy1995particle}.

Learning-based inverse models instead move simulator evaluations to offline
training. They can estimate parameters directly or initialize later optimization
\cite{gabrielli2019multi,han2024learning}. We use the direct variant: a compact
tree ensemble maps fixed-dimensional response descriptors to the six targets in
one pass, and the official simulator then regenerates the corresponding response.
We evaluate parameter accuracy on labeled synthetic data and forward consistency
on the released responses, whose physical parameters are hidden.

\section{Method}

\begin{figure*}[t!]
\centering
\includegraphics[width=0.92\textwidth]{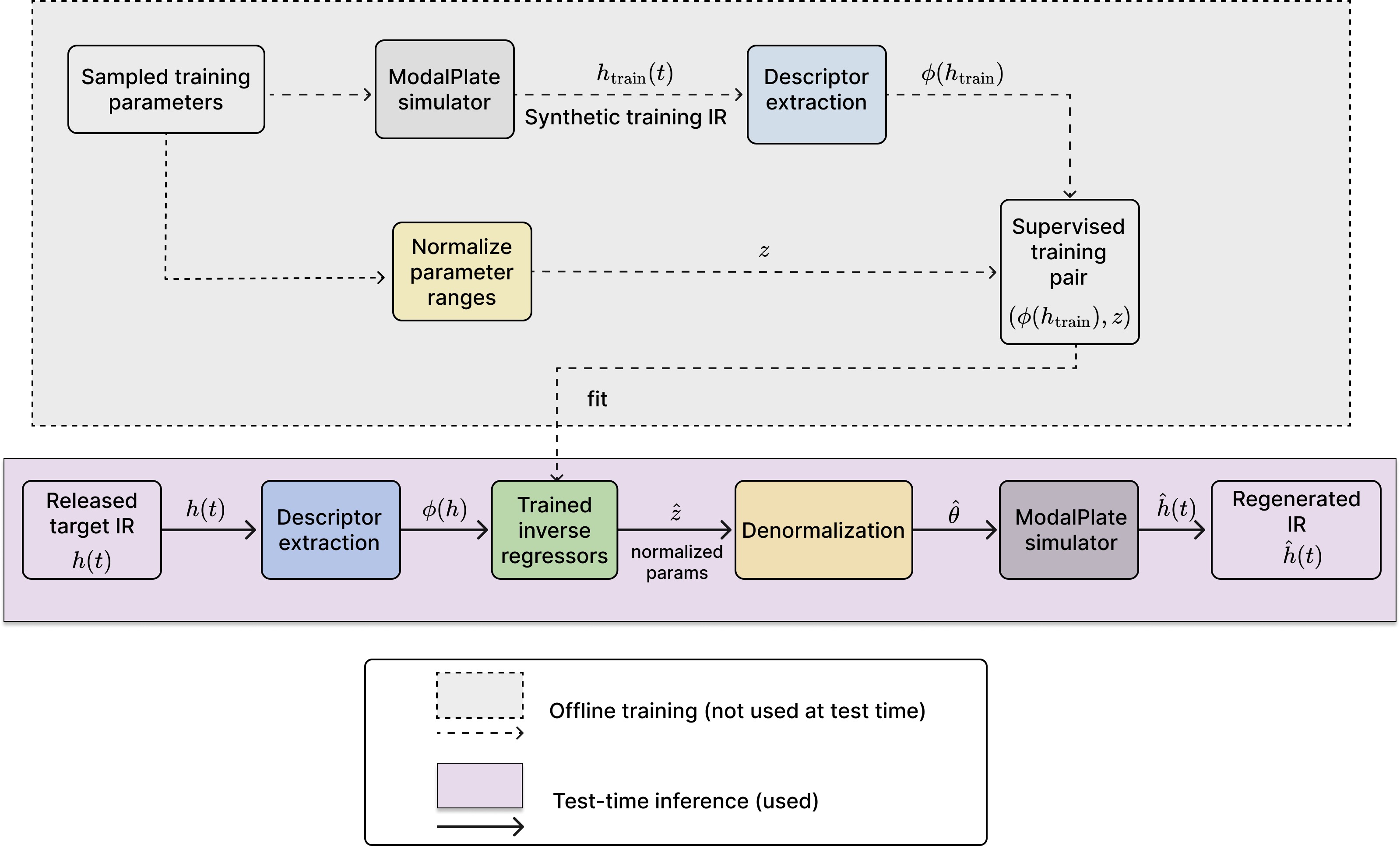}
\caption{Overview of the Task A pipeline. Solid arrows show prediction for a
released challenge response. Dashed gray arrows show how simulated training
examples are made and used to fit the regression model.}
\label{fig:taska-pipeline}
\end{figure*}

Figure~\ref{fig:taska-pipeline} summarizes the offline training and test-time
paths. The complete inference-and-regeneration pipeline is
\begin{equation}
h(t) \rightarrow \phi(h) \rightarrow \hat z \rightarrow \hat\theta
\rightarrow \hat h(t),
\end{equation}
where $h(t)$ is the target response, $\phi(h)$ is its feature vector, $\hat z$ is
the normalized estimate, $\hat\theta$ is the estimate in physical units, and
$\hat h(t)$ is the regenerated response. Simulator calls are used for offline
training and final regeneration. Estimation itself requires only feature
extraction and regression. Training pairs are produced by sampling raw
parameters, synthesizing responses, and extracting $\phi(h)$. Separate labeled
simulations are used for validation because the released labels are hidden.

\subsection{Simulator-based data generation}

Training data are generated with the public ModalPlate simulator
\footnote{\url{https://github.com/LOGUNIVPM/1st-DAFx-Challenge/tree/main/ModalPlate}.}
Variable raw simulator parameters are sampled from
the public ranges, with $L_y\in[1.1,4.0]$, $h\in[10^{-3},5\cdot10^{-3}]$,
$T_0\in[10^{-2},10^3]$, $\rho\in[2430,21230]$,
$E\in[6.7\cdot10^{10},2.2\cdot10^{11}]$, and
$x_o,y_o\in[0.51,1.0]$. We synthesize 5 s displacement impulse responses
matching the released target duration. The corresponding identifiable
parameters $\theta$ are retained as supervised labels. The final estimator is
trained on 1000 synthetic examples. We use two independently generated labeled
sets that are not included in training. \emph{Validation 1} contains 24
responses and was used during model development. \emph{Validation 2} contains
12 responses and was generated only after the final estimator was fixed,
providing a fresh check against dependence on one favorable validation set.

\subsection{Response feature extraction}

Each impulse response is mapped to a fixed-length feature vector designed to
summarize response scale, spectral shape, and decay behavior. The final feature
vector contains 372 scalar values: global time-domain and energy summaries,
early-to-late energy ratios, full-response spectral centroid and tilt, statistics
from nine broad spectral bands, group-delay summaries over four bands, and six
broad frequency-region summaries with band-limited decay statistics. These
features are computed directly from the unnormalized response, without using
target metadata.

\subsection{Predicting normalized parameters}

The six target variables have different scales, so each target is first mapped
to its normalized position inside its allowed range,
\begin{equation}
z_i = \frac{\theta_i-\theta_i^{\min}}
{\theta_i^{\max}-\theta_i^{\min}}.
\end{equation}
The derived ranges induced by the raw simulator bounds are
$\mu\in[2.43,106.15]$, $D/\mu\in[0.28,201.2]$, and
$T_0/\mu\in[9.4\cdot10^{-5},411.5]$. For $L_y$ and $(x_o,y_o)$, which are
already target variables in the simulator, we use their public bounds directly.
These ranges follow from the public raw-parameter bounds using
$\mu=\rho h$, $D=Eh^3/[12(1-\nu^2)]$, and therefore
$D/\mu=Eh^2/[12(1-\nu^2)\rho]$ and $T_0/\mu=T_0/(\rho h)$, where the
Poisson ratio is fixed at $\nu=0.25$.

The final estimate averages two regressors trained on the same
feature-target pairs, namely an ExtraTrees (ET) regressor
\cite{geurts2006extremely} and a histogram gradient-boosting (HGB) regressor
\cite{friedman2001greedy}. We train one model for each of the six normalized
parameters. The ET model uses 500 trees, minimum leaf size 2, and robust
feature scaling. The HGB model uses 250 boosting iterations, learning rate 0.04,
$\ell_2$ regularization $10^{-3}$, and standard feature scaling. For a target
response,
\begin{align}
\hat z_r &= F_r(\phi(h)),
\quad r\in\{\mathrm{ET},\mathrm{HGB}\},\\
\hat\theta_r &= C_\theta\left(\theta^{\min}
+ \hat z_r\odot\Delta\theta\right),\\
\hat\theta &= C_\theta\left(\frac{\hat\theta_{\mathrm{ET}}
+\hat\theta_{\mathrm{HGB}}}{2}\right),
\end{align}
where $F_r$ denotes the six per-parameter regressors in model family $r$,
$\Delta\theta=\theta^{\max}-\theta^{\min}$, $C_\theta$ clips each component to
its allowed physical range, and $\odot$ denotes element-wise multiplication.

\subsection{Forward regeneration for submission}

The challenge also requires an impulse response generated from each estimate.
Because the simulator accepts raw parameters such as $(\rho,h,E,T_0)$ rather
than $(\mu,D/\mu,T_0/\mu)$, we choose a feasible raw tuple matching the estimated
invariants and synthesize its response. This is forward regeneration, not a
separate audio-matching optimization.

\section{Experiments}

Parameter accuracy is evaluated on Validation 1 and Validation 2, since the
physical parameters of the released targets are hidden. For a set of $N$
responses, we report
\begin{equation}
\mathrm{NMSE}=
\frac{1}{6N}\sum_{n=1}^{N}\sum_{i=1}^{6}
\left(
\frac{\hat\theta_{n,i}-\theta_{n,i}}
{\theta_i^{\max}-\theta_i^{\min}}
\right)^2,
\end{equation}
where lower is better. We compare the mean of the 1000 training targets, an
earlier raw multi-output ExtraTrees model, normalized per-parameter ExtraTrees
and HGB models, and their final ensemble. The earlier ExtraTrees candidate uses
246 generic descriptors rather than the final 372-feature set, so it is a
historical system baseline rather than a controlled normalization ablation.
On Validation 2, we additionally run the official PSO baseline with its public
default configuration: 10 particles, five iterations, $(w,c_1,c_2)=(0.2,2,1)$,
and multi-scale spectral loss with STFT configurations $(512,128)$,
$(2048,512)$, and $(8192,2048)$. We evaluate one fixed-seed realization of this
stochastic baseline.

We separately assess forward-response consistency on the 16 released targets.
Using the unnormalized target and regenerated responses, we report absolute
RMS-level error in dB. We also report log-spectrum RMSE up to 12 kHz after
independently peak-normalizing both spectra and applying an $-80$ dB floor.
Temporal energy distribution is measured by
$|\tau(\hat h)-\tau(h)|/\tau(h)$, where the energy-time centroid is
\begin{equation}
\tau(h)=\frac{\sum_n n h[n]^2}{f_s\sum_n h[n]^2}.
\end{equation}
These are local audio-side diagnostics, not official parameter scores.

All runtimes were measured on a MacBook Pro with an Apple M1 Pro 10-core CPU and
16 GB unified memory under macOS 14.5, without an external CUDA GPU. Timing for
the learned estimator excludes synthetic-data generation and training. It
processed all 16 released responses in \SI{15.03}{s}. Since inference uses one feature-
extraction and regression pass, the reported number of optimization
iterations/trials is zero.

\section{Results}

Table~\ref{tab:taska-val} compares parameter accuracy for the final ensemble,
its component models, and the baselines.

\begin{table}[ht]
\caption{NMSE on two independent synthetic sets. Lower is better. Bold marks the
column minimum, and ``--'' denotes an unevaluated configuration.}
\centering
\begin{tabular}{lcc}
\toprule
Model & Validation 1 & Validation 2 \\
\midrule
Training-set mean & 0.047229 & 0.047080 \\
Official PSO (default) & -- & 0.046023 \\
Raw ExtraTrees candidate & 0.026776 & 0.022619 \\
Normalized ExtraTrees & 0.012163 & 0.014946 \\
Normalized HGB & 0.012426 & \textbf{0.012063} \\
Final tree ensemble & \textbf{0.011886} & 0.012935 \\
\bottomrule
\end{tabular}
\label{tab:taska-val}
\end{table}

The normalized per-parameter models outperform the training-set mean and raw
multi-output baseline on both sets. On Validation 2, the final ensemble also reduces
NMSE by 71.9\% relative to the official default PSO run. The ensemble is best on
Validation 1, whereas HGB is slightly better on Validation 2, so averaging is not
uniformly beneficial.

The PSO run required \SI{2252.59}{s} for Validation 2, compared with \SI{10.75}{s}
for the final ensemble, including model loading and feature extraction. This is
a 210-fold reduction in parameter-estimation time at the tested default PSO
budget.

Table~\ref{tab:taska-per-param} breaks down the final ensemble error by
parameter. The normalized stiffness and tension terms have the lowest errors on
both sets. The largest error is associated with an output-position coordinate,
although the harder coordinate differs between sets.

\begin{table}[ht]
\caption{Final ensemble per-parameter NMSE. Lower is better. Bold marks the
column minimum.}
\centering
\begin{tabular}{lcc}
\toprule
Parameter & Validation 1 & Validation 2 \\
\midrule
$\mu$ & 0.009007 & 0.009952 \\
$D/\mu$ & \textbf{0.000445} & \textbf{0.000059} \\
$T_0/\mu$ & 0.000910 & 0.000732 \\
$L_y$ & 0.016485 & 0.021350 \\
$x_o$ & 0.032235 & 0.022773 \\
$y_o$ & 0.012235 & 0.022743 \\
\bottomrule
\end{tabular}
\label{tab:taska-per-param}
\end{table}

Table~\ref{tab:taska-audio} summarizes forward-response consistency over all 16
released targets. IR 0012 is the largest mismatch for all three diagnostics.
Four illustrative listening, waveform, and spectrogram comparisons are
available at \taskADemoLink.

\begin{table}[ht]
\caption{Audio-side diagnostics on the 16 released targets.}
\centering
\begin{tabular}{lcc}
\toprule
Diagnostic & Median & Largest mismatch \\
\midrule
Absolute RMS-level error & \SI{1.04}{dB} & \SI{18.56}{dB}, IR 0012 \\
Log-spectrum RMSE & \SI{4.63}{dB} & \SI{13.66}{dB}, IR 0012 \\
Energy-time error & 1.23\% & 25.19\%, IR 0012 \\
\bottomrule
\end{tabular}
\label{tab:taska-audio}
\end{table}

\section{Discussion}

The similar ensemble NMSE on the two independent synthetic sets supports
repeatability within the matched simulator distribution rather than dependence
on one favorable split. The much smaller errors for normalized stiffness and
tension than for output position suggest that global modal-frequency patterns
are easier to identify than mode-visibility patterns from one observation point.
The reversal between HGB and the ensemble across sets also cautions against
claiming a uniform benefit from model averaging.

The released-target diagnostics show that forward-response consistency varies
across targets, but they cannot establish parameter accuracy because the labels
are hidden. Parameter error and response similarity are related but not
interchangeable: small parameter errors can shift modal frequencies and
accumulate phase differences, whereas different estimates may retain similar
spectral or temporal characteristics. The audio diagnostics therefore
complement rather than replace the synthetic parameter evaluation.

The official PSO comparison shows that amortizing simulator calls across offline
training can outperform a low-budget per-target search at much lower test-time
cost, consistent with learned inverse estimation
\cite{gabrielli2019multi,han2024learning}. It is nevertheless one stochastic
realization. Larger swarms, more iterations, or restarts may improve PSO,
while our timing excludes data generation and training. We also did not compare
with differentiable refinement \cite{jin2024diffsound,diaz2025fast}. Both
validation sets are small and simulator-matched, the historical baseline is not
a controlled ablation, and the estimator provides no uncertainty. Future work
should isolate each design choice and study uncertainty-aware refinement,
mismatched or measured plates, and multiple observation locations.

\section{Conclusion}

We presented a fast, simulation-trained Task A estimator. On two independent
synthetic sets, the normalized tree models outperform the local baselines. On
a common synthetic validation set, the final ensemble also surpasses an
official default PSO run at much lower inference cost. The pipeline
provides a reproducible baseline for single-response plate-reverberation
inversion and could serve as a fast initializer for uncertainty-aware physical
refinement. Uneven parameter errors and released-target consistency also
motivate evaluation on measured plates and with multiple observation points.

\bibliographystyle{IEEEtranDAFx}
\bibliography{references}

\end{document}